\documentclass[12pt]{iopart}
\usepackage{iopams}
\usepackage{graphicx}

\begin{document}

\title[Periodic orbit bifurcations as an ionization mechanism]{Periodic orbit bifurcations as an ionization mechanism: The bichromatically driven hydrogen atom}

\author{S. Huang$^1$, C. Chandre$^2$, T. Uzer$^1$}

\address{$^1$ Center for Nonlinear Science, School of Physics,
Georgia Institute of Technology, Atlanta, Georgia 30332-0430,
U.S.A.}
\address {$^2$ Centre de Physique Th\'eorique\footnote{UMR 6207 of the CNRS, Aix-Marseille and Sud Toulon-Var Universities. Affiliated with the CNRS Research Federation FRUMAM (FR 2291). CEA registered research laboratory LRC DSM-06-35.} - CNRS, Luminy - Case 907, 13288 Marseille cedex 09, France}

\ead{gtg098n@mail.gatech.edu}


\begin{abstract}
We investigate the multiphoton ionization of hydrogen driven by a
strong bichromatic microwave field. In a regime where classical
and quantum simulations agree, periodic orbit analysis captures
the mechanism~: Through the linear stability of periodic orbits we
match qualitatively the variation of experimental ionization rates
with control parameters such as the amplitudes of the two modes of
the field or their relative phases. Moreover, we discuss an
empirical formula which reproduces quantum simulations to a high
degree of accuracy. This quantitative agreement shows the
mechanism by which short periodic orbits organize the dynamics in
multiphoton ionization. We also analyze the effect of longer pulse
durations. Finally we compare our results with those based on the
peak amplitude rule. Both qualitative and quantitative analyses
are implemented for different mode locked fields. In parameter
space, the localization of the period doubling and halving allows
one to predict the set of parameters (amplitudes and phase lag)
where ionization occurs.

\end{abstract}


\submitto{\JPB}

\maketitle

\section{INTRODUCTION}
\label{sec1}

Among systems at the atomic level, one-electron systems are the
most fundamental and have proven to be a source of numerous
surprising discoveries \cite{blumel9}~: The multiphoton ionization
of hydrogen in a strong microwave field \cite{Bayfield} is such a
simple system with surprisingly complex dynamics. With the
development of theory of chaos, its stochastic and diffusional
nature was elucidated \cite{Meerson}, and intense research
activity in the last three decades has resulted in a rather
complete understanding of this problem
\cite{Casati5,Jenson,kochreview}. In the recent years, the
attention has shifted from understanding the physical process to
manipulating or controlling it~\cite{rabitz,shapiro}. In this
context control refers to tailoring the physical behavior of
dynamical systems (which generically show chaotic dynamics) using
``knobs'' (i.e., suitable external parameters). Identifying such
knobs and understanding the mechanism by which they affect the
dynamics is the ultimate goal of such research.

Bichromatic pulses \cite{ko,Ehlotzky} are natural tools in atomic
control research because they offer practical control parameters
such as polarization, amplitudes and phases
\cite{Howard,Haffmans,ivanov,Buchleitner,petrosyan,Sirko1,Sirko,batista,PMKoch,rangan,greeks}.
It has been shown that it is possible to use the direction of
transport in a ratchet by varying the phase lag in the bichromatic
pulse \cite{Carlo}. In this manuscript we consider the ionization
behavior of a one-dimensional hydrogen atom driven by a strong
bichromatic linearly polarized electric field which is modeled by
the following one-dimensional Hamiltonian $(x>0)$ in atomic units
\begin{equation}
\label{Hatom} H=\frac{p^{2}}{2}-\frac{1}{x}+F_{h} x\sin(h\omega
t)+F_{l} x\sin (l\omega t+\phi),
\end{equation}
where the indices $l$ and $h$ refer to the low and high frequency
modes with frequencies $l\omega$ and $h\omega$, respectively.
These two modes are frequency locked (and denoted $h$:$l$), i.e.\
$l$ and $h$ are integers. The experimental results on ionization
probability obtained in Refs.~\cite{Sirko,PMKoch} for two cases,
mode lockings 3:1 and 3:2, show two very distinct regimes as the
phase lag $\phi$ is varied~: For 3:1, the ionization probability
shows a plateau in phase located at a rather high value of
ionization probability, and for 3:2, the ionization probability
shows no such plateau but rather a small value of ionization
probability. These results were confirmed by quantum calculations
on the one-dimensional model~(\ref{Hatom}).

The purpose of this manuscript is to show that these experimental
observations are qualitatively and quantitatively captured using a
periodic orbit analysis, which reveals the classical bifurcations
responsible for ionization. Our analysis also allows the
prediction of ionization at other values of parameters without
resorting to large numerical simulations. A part of the results
presented in this manuscript were announced in a Fast Track
Communication~\cite{shuang2}.

This paper is organized as follows: First, in Sec.~\ref{sec2}, we
summarize the residue analysis of periodic
orbits~\cite{Bachelard}. In Sec.~\ref{sec3} and Sec.~\ref{sec4},
we consider two cases, the 3:1 mode locking and 3:2 one,
respectively. These two cases show drastically different
ionization behavior which we explain through bifurcations of
selected periodic orbits. In these two cases, we described the
bifurcations (if any) in the system. We also compute the
bifurcation surface which is defined by the set of parameters
where a change of linear stability (associated with a period
doubling or halving) has occurred in the chaotic sea. In
Sec.~\ref{sec5} we discuss the generalities of residue curve
behavior for other mode lockings.

\section{Residue analysis of periodic orbits}
\label{sec2}

First, we map Hamiltonian~(\ref{Hatom}) into action-angle
variables (of the unperturbed system $F_h=F_l=0$) such that the
principal quantum number $n$ is reflected by action $J$. The
relationship between $J$ and $n$ reads
$$
J=(t_{0} \omega)^{1/3}n,
$$
where $t_{0}=2.4188843243\times10^{-17} s$ (atomic time unit) and
$\omega = 2\pi f =12\pi\mbox{ GHz}$ in both cases. The
action-angle variables~\cite{Leopold} are denoted $(J,\theta)$ and
obtained through the canonical change of coordinates
$x=2J^{2}\sin^{2}\varphi$, $p=J^{-1}\cot\varphi$ with
$\theta=2\varphi-\sin2\varphi$.

We assume $\omega=1$ without loss of generality (after a rescaling
$H'=\omega^{-2/3} H$, $t'=\omega t$, $x'=\omega^{2/3}x$,
$p'=\omega^{-1/3}p$ and consequently $\phi'=\phi$ and
$F'=\omega^{-4/3}F$). The Hamiltonian~(\ref{Hatom})
becomes~\cite{Casati5}
\begin{eqnarray}
H&=&-\frac{1}{2J^{2}}+2J^{2}[F_{h}\sin(ht)\nonumber
\\ &&+F_{l}\sin (l t+\phi)]\left( a_{0}/2+\sum_{k=1}^\infty{a_{k}\cos
k\theta}\right),\label{HatomAA}
\end{eqnarray}
where $a_{n}=[J_{n}(n)-J_{n-1}(n)]/n$ and $J_{n}$'s are Bessel
functions of the first kind. Note that for a given mode locking
$h$:$l$, there are three variables $(J,\theta,t)$ and three
parameters $(F_h,F_l,\phi)$. We denote this Hamiltonian
$H(J,\theta,t;F_h,F_l,\phi)$. In general, for a given set of
parameter values, the phase space as depicted on a Poincar\'e
section (a stroboscopic plot with period $2\pi$, see
Fig.~\ref{fig:fig1}), is composed of a mixture of regular
structures surrounded by a chaotic sea. More precisely, the lower
part of phase space ($J$ small) is composed of rotational
invariant tori (which persist from the integrable case $F_h=F_l=0$
as asserted by KAM theorem). The upper part is formed by regular
islands surrounded by an unbounded chaotic sea. The ionizing
trajectories are the ones in the chaotic sea which are unbounded
($J$, or equivalently $n$, becomes progressively large). At the
center of the regular islands, there are elliptic (stable)
periodic orbits which organize the regular motion around them.
These periodic orbits partly result from the break-up of resonant
tori into a pair of elliptic/hyperbolic orbits (according to
Birkhoff's theorem). Other periodic orbits result from the
bifurcation of these orbits.

The general idea which is applied here is to follow a finite set
of periodic orbits (both elliptic or hyperbolic) which have been
identified as important. The criteria of choice combine several
factors: the size of the island, the period and the location. For
each periodic orbit of this set, we compute its location
$(J,\theta)$ and its linear stability property as given by the
residue (which is to be defined below). As the three parameters
are varied, we follow the locations and residues of these orbits
instead of computing the Poincar\'e section for each value of
parameters. This allows us to have a clear idea of what is going
on in phase space and to predict ionization thresholds.

In what follows, we perform two kinds of computations~: First, we
compute the residue curves which are obtained as functions of the
relative phase $\phi$ for fixed values of the amplitudes
$(F_h,F_l)$. Second, we compute bifurcation surfaces which are
defined as the set of parameters $(F_h,F_l,\phi)$ where a change
of linear stability has occurred.

In order to start monitoring the stability of a family of periodic
orbits, we first consider a specific periodic orbit, denoted
${\mathcal O}(0)$, of Hamiltonian~(\ref{HatomAA}) for $\phi=0$
which is our reference case. Numerically it is determined using a
modified Newton-Raphson multi-shooting algorithm as described in
Ref.~\cite{chaosbook}. The initial condition for launching the
iterative algorithm can be taken from a Poincar\'e section, for
instance. As $\phi$ is continuously varied, the
orbit ${\mathcal O}(0)$ deforms continuously into ${\mathcal
O}(\phi)$, whose period is denoted $T(\phi)$. In addition to its
location, we also monitor its linear stability properties
given by the integration of the reduced tangent flow
$$
\frac{d{\mathcal J}_\phi^t}{dt}={\mathbb J}\nabla^2
H(J,\theta,t;\phi) {\mathcal J}^t_\phi,
$$
where ${\mathbb J}=\left(\begin{array}{cc} 0 & 1\\ -1 & 0
\end{array}\right)$ and $\nabla^2 H$ is the two-dimensional
Hessian matrix (composed of second derivatives of $H$ with respect
to its canonical variables $J$ and $\theta$). The initial
condition is ${\mathcal J}_\phi^0={\mathbb I}_2$ (the
two-dimensional identity matrix). The two eigenvalues of the
monodromy matrix ${\mathcal J}_\phi^{T(\phi)}$ which make a pair
$(\lambda(\phi),1/\lambda(\phi))$ determine the stability properties (the flow
is volume preserving, thus the determinant of ${\mathcal
J}_\phi^{T(\phi)}$ is equal to 1). The periodic orbit is elliptic
if the spectrum is $({\mathrm e}^{i\omega(\phi)},{\mathrm
e}^{-i\omega(\phi)})$ (stable, except for some particular cases),
or hyperbolic if the spectrum is $(\lambda(\phi),1/\lambda(\phi))$
with $\lambda(\phi)\in{\mathbb R}^*$ (unstable). The stability properties can be deduced in a concise form using Greene's residue $R$~\cite{gree79,mack92}
$$
R(\phi)=\frac{2-\mbox{tr}{\mathcal J}_\phi^{T(\phi)}}{4}.
$$
If $R(\phi)\in ]0,1[$, the periodic orbit is elliptic; if
$R(\phi)<0$ or $R(\phi)>1$ it is hyperbolic; and if $R(\phi)=0$
and $R(\phi)=1$, it is parabolic. Generically, periodic orbits and
their linear stabilities are robust against small changes of
parameters, except at specific values where bifurcations occur
\cite{Cary}. These rare events affect the dynamical behavior
drastically. They can be associated with an enhancement as well as
a reduction of stability depending on the type of bifurcations, as
shown in Refs.~\cite{Bachelard,Cary}. We identify the bifurcations
(if any) of a set of short periodic orbits, i.e., the type and the
value of the parameter $\phi_c$ where they bifurcate. This
provides a way to foretell if a relatively high ionization rate
should be expected or not. The importance of considering two
associated Birkhoff periodic orbits (i.e.\ periodic orbits with
the same action but different angles in the integrable case, one
elliptic and one hyperbolic), was emphasized in
Ref.~\cite{Bachelard}. The main reason is that it allows one to
discard some specific bifurcations (like collisions or exchanges
of stability) and draw appropriate conclusions concerning the
enhancement or reduction of stability.

In the following sections, we analyze the short-time dynamics
through residues of selected periodic orbits. We compute residue
curves $\phi \mapsto R(\phi,F_h,F_l)$ for fixed values of the
amplitudes $F_h$ and $F_l$, and bifurcation surfaces associated
with a given periodic orbit defined as the set of parameters such
that $R(\phi,F_h,F_l)=1$. We correlate these results with
ionization probabilities obtained experimentally and also by
quantum simulations through an empirical formula.

\section{3:1 mode locking}
\label{sec3}

For this mode locking, we consider $F_h=24\mbox{ Vcm}^{-1}$ and
$F_l=53.4\mbox{ Vcm}^{-1}$ as in Ref.~\cite{PMKoch}. These values
correspond to the dimensionless values
$F_h=24v_{0}(t_{0}\omega)^{-4/3}/d_0=0.5278$ and $F_l=53.4
v_{0}(t_{0}\omega)^{-4/3}/d_0=1.1743$, where $v_{0}=0.036749326$
and $d_{0}=1.88972613\times 10^{8}$ are both conversion factors,
$t_{0}=2.4188843243\times 10^{-17} s$ (atomic time unit). These
values are referred as Case $(I)$ in what follows.

\subsection{Poincar\'e section}
\label{sec3A}

\begin{figure}
 \centering
 \includegraphics[width=8.5cm,height=7.5cm]{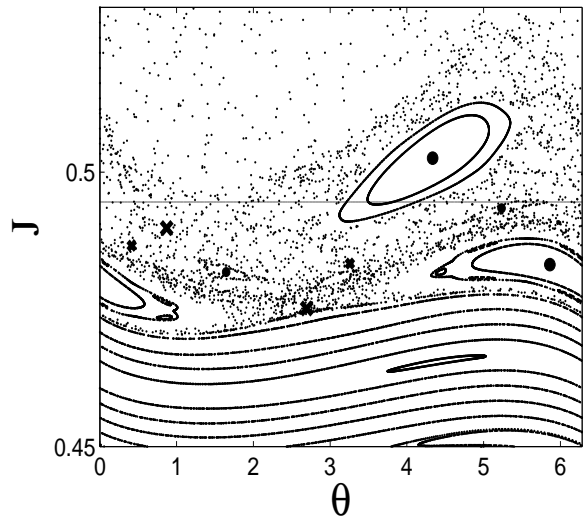}
\caption{\label{fig:fig1}Poincar\'e section of
Hamiltonian~(\ref{HatomAA}) for Case $(I)$ at $\phi=0$. Full big
circles (or big crosses, respectively ) indicate the two elliptic
(resp.\ hyperbolic) periodic orbits with period $2\pi$ we
consider. Full small circles ( or small crosses, respectively)
indicate the two elliptic (resp.\ hyperbolic) periodic orbits with
period $4\pi$ we consider. The horizontal line corresponds to the
principal quantum number $n=51$.}
\end{figure}

Figure~\ref{fig:fig1} shows a Poincar\'e section of
Hamiltonian~(\ref{HatomAA}) for Case $(I)$ at $\phi=0$. The phase
space is divided into two main parts~: a lower regular containing
many invariant tori, and a upper chaotic sea where trajectories
escape rapidly to unbounded actions $J$ (ionized trajectories). We
notice two main islands in the chaotic sea. At the centers of
these islands sit elliptic periodic orbits with period $2\pi$
(indicated by full circles). In addition, there is also a period 2
island in between these two main islands (associated with an
elliptic and hyperbolic periodic orbits with period $4\pi$). In
the region of phase space around $J_i\approx 0.49$ (indicated by a
straight line) where the initial states $n=51$ are
prepared~\cite{PMKoch}, any regular structures like the main
islands and smaller ones are associated with trappings and hence
reduce the ionization rate.

\begin{figure}
 \centering
 \includegraphics[width=8.5cm,height=7.5cm]{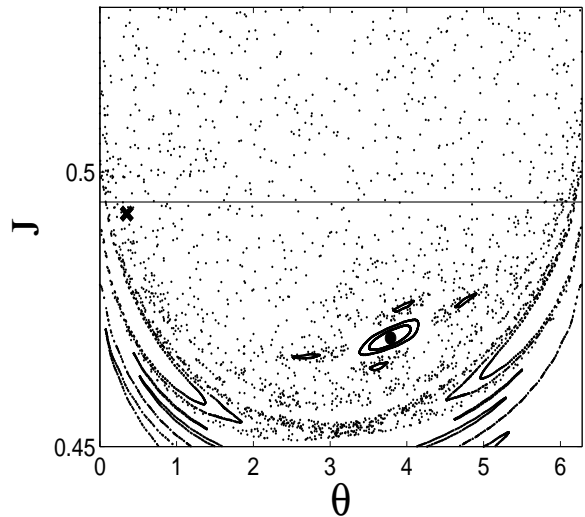}
\caption{\label{fig:fig2}Poincar\'e section of
Hamiltonian~(\ref{HatomAA}) for Case $(I)$ at $\phi=\pi/3$. Full
circle (respectively cross) indicates the elliptic (resp.\
hyperbolic) periodic orbit with period $2\pi$ we consider. The
horizontal line corresponds to the principal quantum number
$n=51$.}
\end{figure}

Figure~\ref{fig:fig2} shows a Poincar\'e section of
Hamiltonian~(\ref{HatomAA}) for Case $(I)$ at $\phi=\pi/3$.
Apparently the upper elliptic periodic orbit of period $2\pi$ in
Fig.~\ref{fig:fig1} has disappeared in chaotic sea. This case is
associated with a higher ionization probability. Therefore, the
loss of stability of the main periodic orbit near $J_i$ plays a
major role in the ionization process.

The residue method will monitor the location and stability of
elliptic periodic orbits as well as their associated hyperbolic
periodic orbits~\cite{Bachelard} as a function of the parameters
of the system.

\begin{figure}
 \centering
 \includegraphics[width=8.5cm,height=7.5cm]{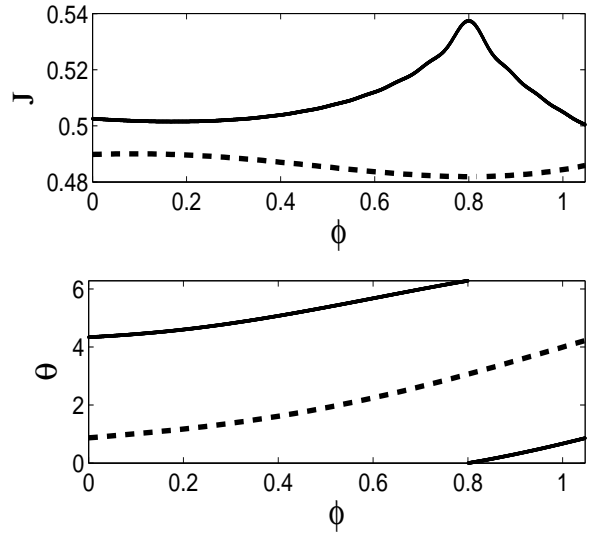}
\caption{\label{fig:fig3}The positions of the upper elliptic and
hyperbolic periodic orbits with period $2\pi$ on the Poincar\'e
section as functions of $\phi$ for Case $(I)$. The solid curves
and the dashed ones correspond to the upper elliptic and
hyperbolic periodic orbits of Fig.~\ref{fig:fig1} respectively.}
\end{figure}

For Case $(I)$, Figure~\ref{fig:fig3} shows the positions of the
upper elliptic and hyperbolic periodic orbits of period $2\pi$ on
the Poincar\'e section as functions of $\phi$. We notice that the
action $J$ is changing weakly as $\phi$ is varied. In contrast,
its angle $\theta$ is very sensitive to this parameter.

\subsection{Residue curve}
\label{sec3B}

\begin{figure}
 \centering
 \includegraphics[width=8.5cm,height=7.5cm]{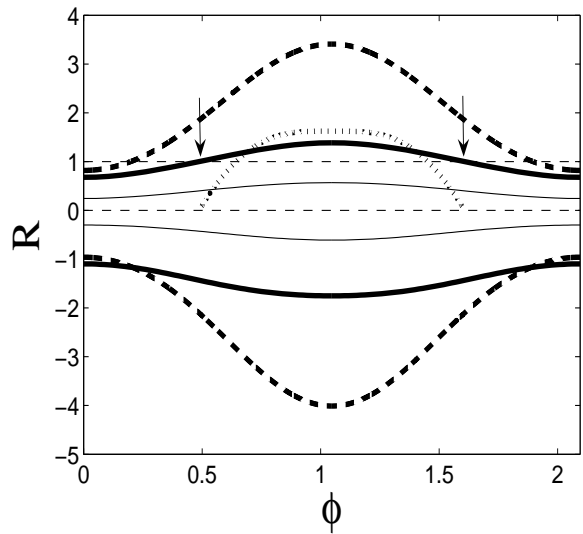}
\caption {\label{fig:fig4} Residue curves for the four periodic
orbits with period $2\pi$ (solid curves) and the two periodic
orbits with period $4\pi$ (dashed bold curves), indicated by
crosses and circles on Fig.~\ref{fig:fig1} for Case $(I)$. The
solid bold curves are for the upper set of elliptic/hyperbolic
orbits of period $2\pi$. Small arrows indicate where bifurcations
happen. The dotted bold curve between two arrows is associated
with the residues of the elliptic periodic orbit with period
$4\pi$ born of the period doubling bifurcation.}
\end{figure}

We follow the residue for each of the elliptic and hyperbolic
periodic orbits mentioned above as the parameter $\phi$ varies for
Case $(I)$. Figure~\ref{fig:fig4} shows the residue curves of
these orbits.

In Fig.~\ref{fig:fig4}, we monitor the upper elliptic periodic
orbit (period $2\pi$) of Fig.~\ref{fig:fig1} from $\phi=0$. This
periodic orbit remains elliptic ($R(\phi)\in ]0,1[$) until
$\phi_c\approx 0.49$ where a bifurcation occurs. At this critical
point, the orbit turns parabolic. Increasing $\phi$ further, it
turns and remains hyperbolic ($R(\phi)> 1$) until $2\pi/3-\phi_c$
where another bifurcation appears, making the orbit elliptic
again. This bifurcation process is of the period doubling kind at
$\phi_c$ and a period halving at $2\pi/3-\phi_c$ (by symmetry).

\begin{figure}
 \centering
 \includegraphics[width=8.5cm,height=7.5cm]{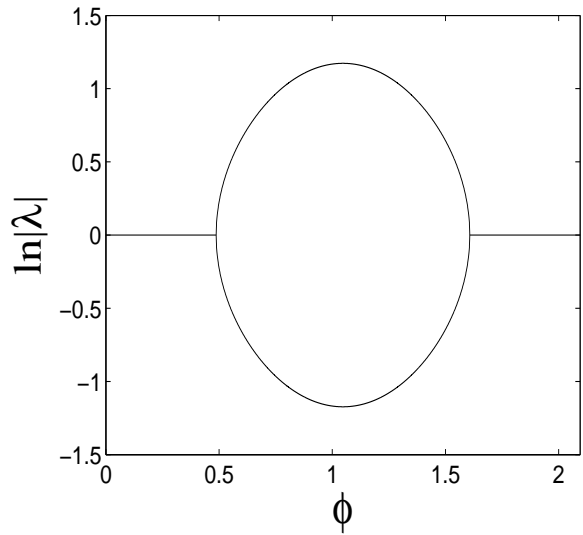}
\caption {\label{fig:fig5} Bifurcation diagram for Case $(I)$
showing the bifurcations indicated by arrows in
Fig.~\ref{fig:fig4}.}
\end{figure}

The bifurcation diagram for Case $(I)$ is shown in
Fig.~\ref{fig:fig5}. The computation of $\log|\lambda_\pm(\phi)|$
(where $\lambda_\pm(\phi)$ are the two eigenvalues of the
monodromy matrix ${\mathcal J}_\phi^{T(\phi)}$ associated with the
upper elliptic periodic orbit) gives
$\log|\lambda_\pm(\phi)|=0$ before the bifurcation and
$\log|\lambda_\pm(\phi)|\propto \pm \sqrt{\phi-\phi_c}$ right after the
bifurcation.

\begin{figure}
 \centering
 \includegraphics[width=8.5cm,height=7.5cm]{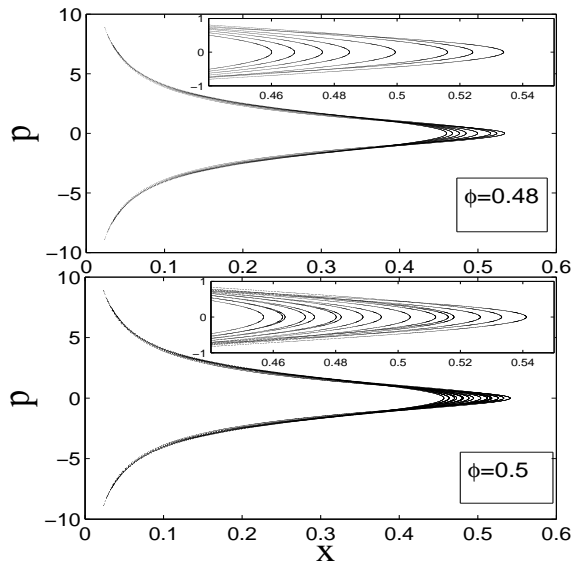}
\caption {\label{fig:fig6} Upper elliptic periodic orbit (period
$2\pi$) of Case $(I)$ undergoing the periodic doubling bifurcation
in $x-p$ representation. The insets shows the doubled number of
branches indicating periodic doubling bifurcation.}
\end{figure}

Figure~\ref{fig:fig6} shows a projection of the upper elliptic
periodic orbit undergoing the period doubling bifurcation in $x-p$
representation where we note the doubling of the number of
branches.

Note that while the upper elliptic periodic orbit undergoes a
bifurcation as $\phi$ is varied, the other three periodic orbits
with period $2\pi$ retain the stability properties they had at
$\phi=0$.

In the parameter range $\phi\in ]\phi_c,2\pi/3-\phi_c[$, the upper
part of phase space does exhibit more chaos as it can be shown on
the Poincar\'e section for $\phi=\pi/3$ (see Fig.~\ref{fig:fig2}).

Since the initial atomic beam is taken in the region with
principal quantum number $n=51$ which corresponds to action
$J_i\approx 0.495$, the ionization rate is expected to be higher
in the regime where there are no big islands in the chaotic sea,
i.e.\ for $\phi\in ]\phi_c,2\pi/3-\phi_c[$. On a finer scale, one
has to take into account the smaller regular structures that are
present in the chaotic sea, like for instance the period-2 island
on the Poincar\'e section (period $4\pi$). A similar period
doubling and halving occur at $\phi_{c,1}\approx 0.199$ and
$2\pi/3-\phi_{c,1}\approx 1.895$ respectively, as can be seen on
Fig.~\ref{fig:fig4} (upper bold dashed curve). We notice that the
residue curve for the period $4\pi$ orbit is higher than the upper
solid curve (for the period $2\pi$ orbit) and the phase region
between period doubling ($\phi_{c,1}$) and period halving
($2\pi/3-\phi_{c,1}$) is much wider than the one based on the
period $2\pi$. This observation indicates that for longer pulse
duration for which longer periodic orbits have to be taken into
account, the period $4\pi$ orbit obtained by a repetition of the
period $2\pi$ orbit is more important (as a limiting factor) for
ionization than the orbit with primary period $4\pi$. As a
by-product, the bifurcation at $\phi_{c,1}$ does not affect
significantly the leading ionization behavior based on period
$2\pi$. Another important periodic orbit with period $4\pi$ is the
one born of the period doubling bifurcation of the upper periodic
orbit at $\phi_c$. This orbit is close to the newly hyperbolic
periodic orbit and is at first elliptic, as seen on
Fig.~\ref{fig:fig4}. It is expected that the associated elliptic
islands slow down ionization. This orbit experiences a period
doubling bifurcation at $\phi_{c,2}\approx 0.643$ and halving
bifurcation at $2\pi/3-\phi_{c,2}\approx 1.451$. In summary, the
shortest periodic orbits (period $2\pi$) always play the most
dominant role on ionizations rates, although with larger pulse
durations longer periodic orbits might be taken into account for a
more detailed analysis.

For values of $\phi$ around $\pi/3$, a plateau is expected in the
ionization probability versus $\phi$. The reason is that a
strongly hyperbolic orbit only influences the ionization time and
not the value of the ionization probability. Of course, this is
true provided that the duration of the maximum pulse envelope is
large enough. In the experiment, this is approximately 15 times
the period of the shortest periodic orbits considered. Roughly
speaking, this means that in the chaotic region all the orbits
ionize (i.e.,\ escape to a value of the action
$J_\mathrm{ion}\gtrsim 1.26$) regardless of the hyperbolicity
degree. In Ref.~\cite{PMKoch}, experimental results as well as
one-dimensional quantum calculations show this plateau. From
quantum calculations, $\phi_c\approx 0.5$ was obtained in
Ref.~\cite{PMKoch} which is in very good agreement with the
parameter value $\phi_c\approx 0.49$ at which the bifurcation of
the upper elliptic periodic orbit (period $2\pi$) occurs. If the
duration of the experiment or simulations is longer (two or three
times longer), then higher-order regular structures (like the
regular island of period $4\pi$ in the Poincar\'e section and the
elliptic periodic orbit with period $4\pi$ born of bifurcations)
will play a role and we expect a similar, but smaller, plateau for
$\phi\in ]\phi_{c,2},2\pi/3-\phi_{c,2}[$.

\begin{figure}
 \centering
 \includegraphics[width=8.5cm,height=7.5cm]{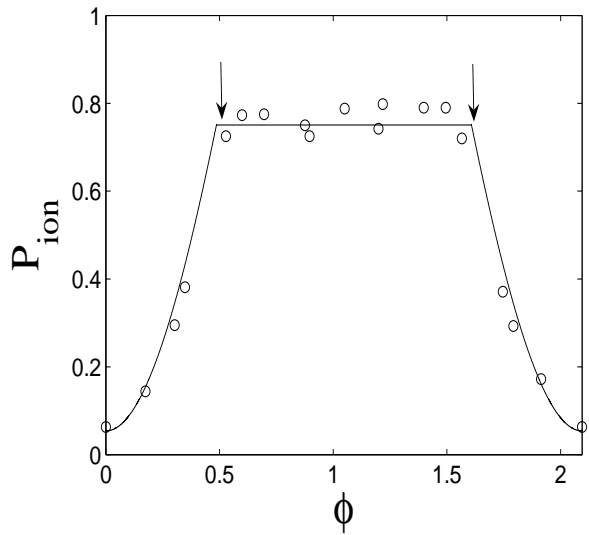}
 \caption{\label{fig:fig7} Normalized ionization probability {\em vs} $\phi$ based on
Eq.~(\ref{Ionempirical}) for Case $(I)$ with $A=-2.52$ and
$B=0.65$. Circles represent the data obtained by one-dimensional
quantum calculations, taken from Ref.~\cite{PMKoch}. Only periodic
orbits with period $2\pi$ are considered.}
\end{figure}

The periodic orbit analysis above elucidates whether or not there
is a significant ionization probability for specific parameter
values, and also where plateaus are expected to occur. This
qualitative agreement highlights the important role played by
these orbits. Furthermore, we can obtain quantitative agreement
concerning the shape of the ionization curve versus phase lag
$\phi$ by using the residue curves. Here we devise an empirical
formula for relative ionization probability in the following way~:
First, the values of $\phi$ giving the highest ionization would be
the ones associated with the highest variations of the residues
(in absolute value) with respect to the minimum ionization.
Second, if the periodic orbit is too far (in action) from the
considered action $J_i$ then it will not influence the dynamics so
there should be a penalizing term depending on its position with
respect to the chosen rescaled action. The relative ionization
probability formula reads~:
\begin{equation}
\label{Ionempirical}
P_{\mathrm{ion}}(\phi)=A+B\sum_{m=1}^{M}\frac{\exp
|R_{m}(\phi)-R_{m}(\phi_{0})|}{\exp
|\overline{J_{m}(\phi)}-J_{i}|},
\end{equation}
where the sum is taken over the $M$ different periodic orbits
considered and
$\overline{J_{m}(\phi)}=\int_{0}^{2\pi}J(\theta)d\theta/2\pi$ is
the action of the periodic orbit $m$. The parameters $A$ and $B$
in Eq.~(\ref{Ionempirical}) are merely a translation and a
dilatation of the curve in order to match the mean value and the
amplitude of variations of $P_{\mathrm{ion}}$ obtained in
Ref.~\cite{PMKoch}. This formula takes into account the value of
the residues at $\phi_{0}$, where the minimum ionization takes
place for each case. The aim is to set up a baseline for each of
the periodic orbits (which is taken here at the value of the
parameter where the ionization is minimal). Specifically for Case
$(I)$ $\phi_{0}=0$~\cite{PMKoch}. In general,
Eq.~(\ref{Ionempirical}) can exhibit values which are greater than
1, which are not relevant. In order to remedy to this problem, we
truncate $P_{\mathrm{ion}}$ at the value where a bifurcation
occurs in accordance with the previous discussion on the relevance
of the degree of hyperbolicity. Therefore in the range where
$R_n(\phi)$ is larger than one, $P_{\mathrm{ion}}$ is constant
(taken as the value of the residue at $\phi_c$ where the
bifurcation occurs).

Figure~\ref{fig:fig7} depicts $P_{\mathrm{ion}}$ given by
Eq.~(\ref{Ionempirical}) versus parameter $\phi$ as well as the
data taken from Ref.~\cite{PMKoch} for Case $(I)$. Only the
shortest periodic orbits with period $2\pi$ are considered as
relevant. We notice that the empirical formula reproduces
accurately the results obtained from quantum calculations. If we
take into account the period $4\pi$ orbits for longer pulse
duration, then a second plateau appears for $\phi\in
]\phi_{c,2},2\pi/3-\phi_{c,2}[$ together with an increase at
$\phi_{c,1}$ and a decrease at $2\pi/3-\phi_{c,1}$.

\subsection{Bifurcation surface}
\label{sec3C}

\begin{figure*}
 \begin{minipage}[t]{8cm}
 \centering
 \includegraphics[width=6.cm,height=5.4cm]{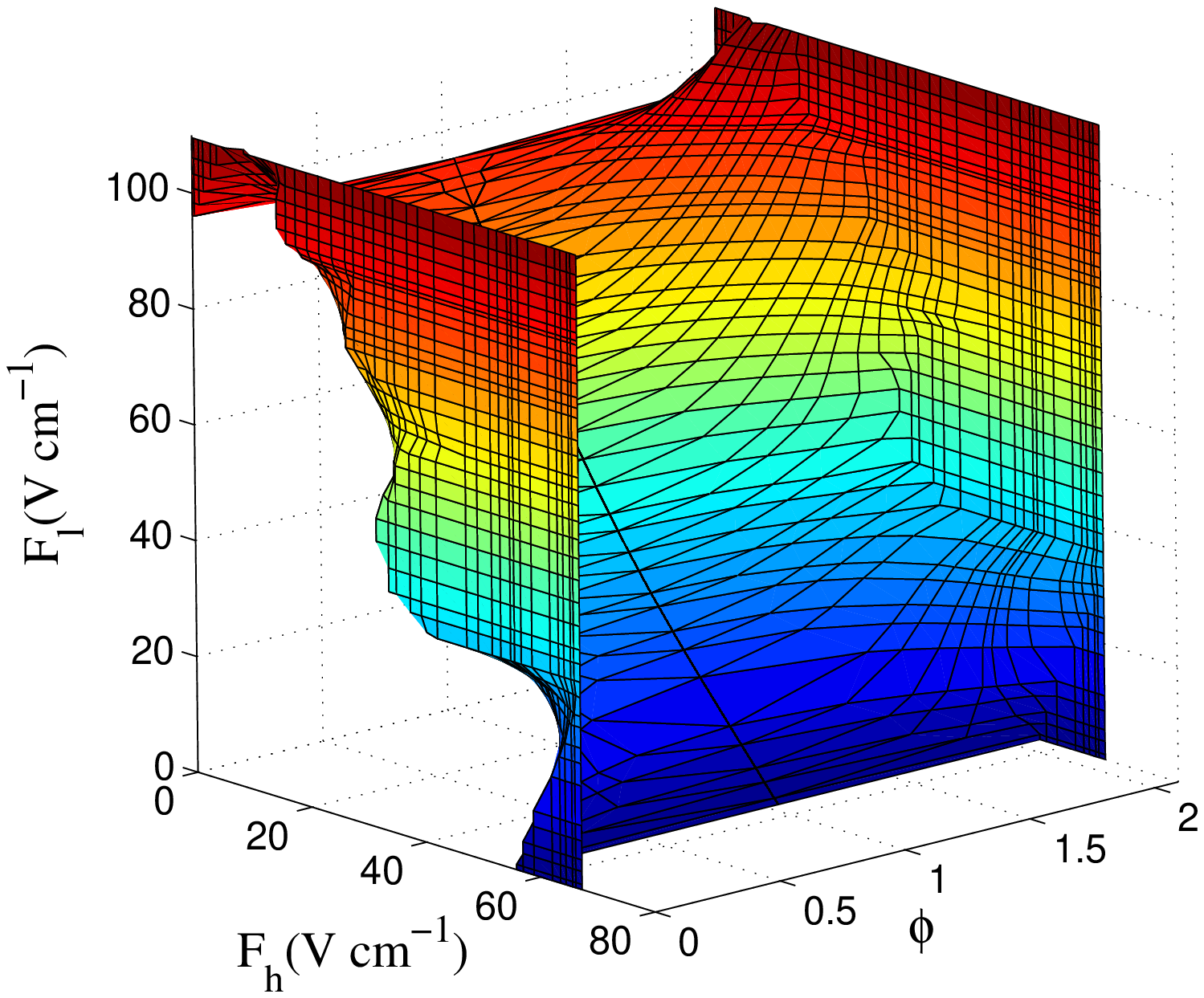}
 \mbox{{\bf (a)} }
 \end{minipage}
 \begin{minipage}[t]{8cm}
 \centering
 \includegraphics[width=6.cm,height=5.4cm]{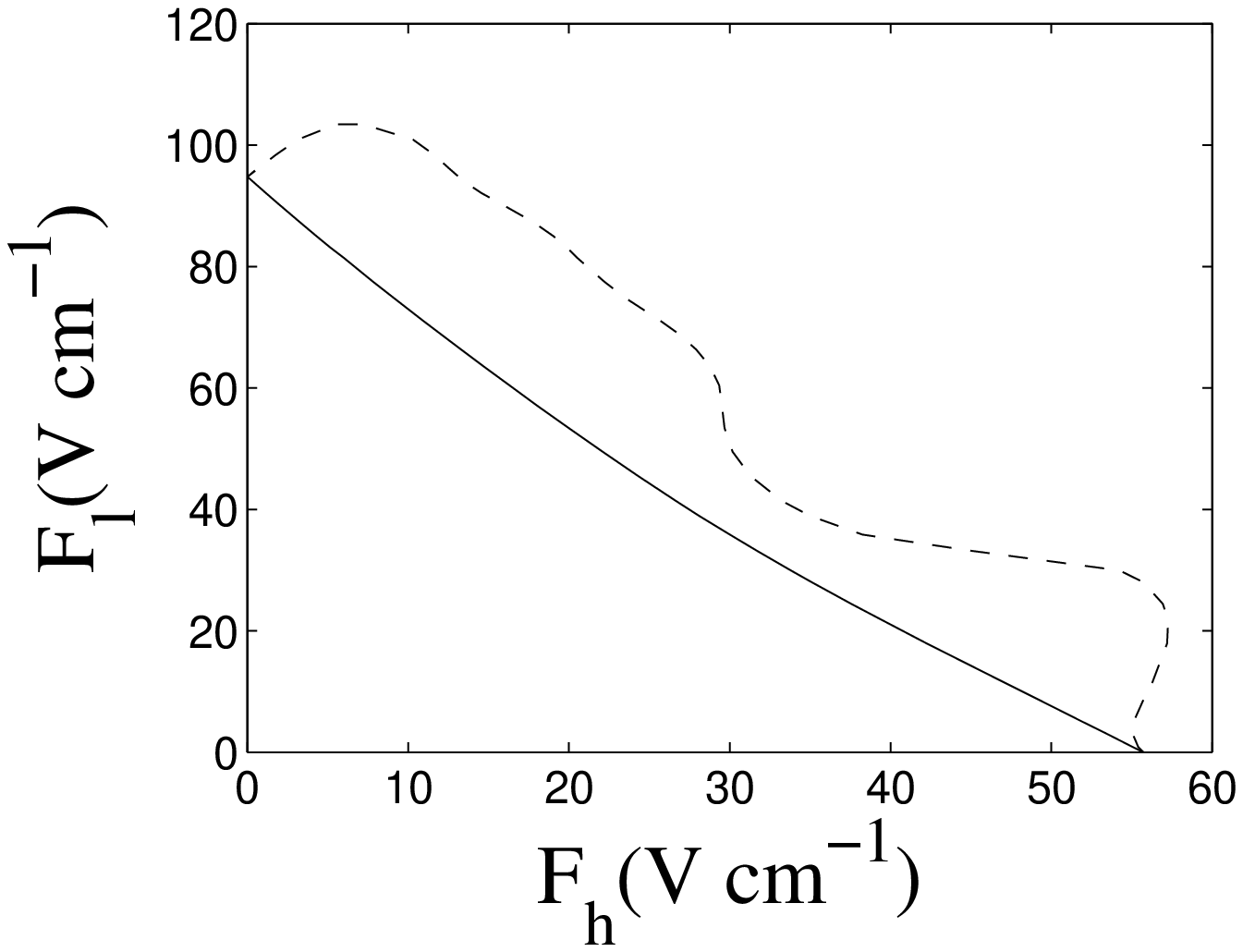}
 \mbox{{\bf (b)} }
 \end{minipage}
\caption{\label{fig:fig8} (color online). (a) Bifurcation surface
in parameter space $(\phi, F_h, F_l)$ for $h$:$l$=3:1. (b) The
continuous curve is a section of the bifurcation surface (a) at
$\phi=\pi/3$, whereas the dashed one is for $\phi=1.95$.}
\end{figure*}

The residue method is carried out to also predict the behavior of
the system as all three parameters (the two amplitudes $F_h$ and
$F_l$ and the phase lag $\phi$) are varied. In
Fig.~\ref{fig:fig8}, we represent the set of parameters where the
upper elliptic periodic orbit (with period $2\pi$) of
Fig.~\ref{fig:fig1} is in fact parabolic (i.e.,\ the set of
parameters where the system undergoes a major bifurcation). The
equation of this surface in parameter space is $R(\phi, F_h,
F_l)=1$. The boundaries of the plateaus in parameter $\phi$ of
Fig.~\ref{fig:fig7} obtained by fixing the two values for $F_h$
and $F_l$ are on this surface. When $F_{h}$ approaches zero, this
surface is less dependent on parameter $\phi$. Table~\ref{tab1}
reports some values based on our analysis which are in good
agreement with experimental results from Ref.~\cite{Sirko}.
\begin{table}
\caption{\label{tab1}Ionization thresholds obtained for
$F_h=6\mbox{ Vcm}^{-1}$, experimentally in Ref.~\cite{Sirko}  and
by the residue method (see Fig.~\ref{fig:fig4}). The 1f case
corresponds to $F_h=0$.}

\begin{indented}
\item[]\begin{tabular}{c|ccc} \br
$F_l (\mbox{ Vcm}^{-1})$ & $\phi=0$ & $\phi=\pi/3$ & 1f\\
\mr
\cite{Sirko} & 107 & 85 & 96\\
\verb"residue" & 109.6 & 81.4 & 94.8\\
\br
\end{tabular}
\end{indented}
\end{table}
Of course, the surface of Fig.~\ref{fig:fig8} could also have been
obtained from tedious numerical simulations of a large number of
classical trajectories for each value of the parameters
$(\phi,F_h,F_l)$. This integration needs to be performed for a
sufficiently long time in order to decide if a given trajectory
leads to ionization or not. In contrast, only one orbit for a
short time (typically the period of the field) is needed for the
residue analysis. Furthermore, using residues, this surface can be
constructed locally without any need to consider all possible
values of the parameters.

\section{3:2 mode locking}
\label{sec4}

In what follows, we study in detail one set of values for the
amplitudes of the fields $F_h=25\mbox{ Vcm}^{-1}$ and
$F_l=33.5\mbox{ Vcm}^{-1}$ which correspond to $F_h=0.5498$ and
$F_l=0.7367$ in dimensionless units. This case will be referred as
Case $(II)$.

\subsection{Poincar\'e section}
\label{sec4A}

\begin{figure}
 \centering
 \includegraphics[width=8.5cm,height=7.5cm]{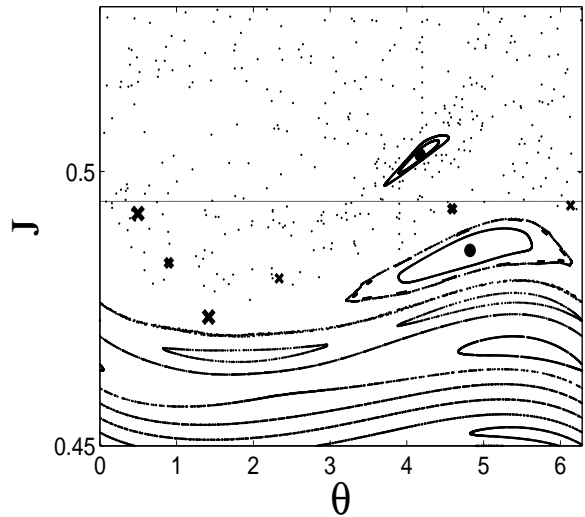}
\caption{\label{fig:fig9}Poincar\'e section of
Hamiltonian~(\ref{HatomAA}) for Case $(II)$ at $\phi=0$. Full
circles (respectively crosses) indicate the two elliptic (resp.\
hyperbolic) periodic orbits with period $2\pi$ we consider. Small
crosses indicate the two hyperbolic periodic orbits with period
$4\pi$ we consider. The horizontal line corresponds to the
principal quantum number $n=51$.}
\end{figure}

Figure~\ref{fig:fig9} shows a Poincar\'e section of
Hamiltonian~(\ref{HatomAA}) for Case $(II)$ at $\phi=0$. Similar
to Case $(I)$, there are two primary islands in the chaotic sea
where two elliptic periodic orbits with period $2\pi$ sit at the
centers, and the two associated hyperbolic orbits are in the
chaotic sea. In addition, there are also two associated hyperbolic
periodic orbits with period $4\pi$ as indicated by small crosses.
One should notice that the (rescaled) principal quantum number
considered in Ref.~\cite{PMKoch} lies in between these two islands
(see the continuous horizontal line). We monitor the stability of
this set of periodic orbits as we have performed for Case $(I)$.

\subsection{Residue curve}
\label{sec4B}

\begin{figure}
\begin{minipage}[t]{8cm}
 \centering
 \includegraphics[width=8.5cm,height=7.5cm]{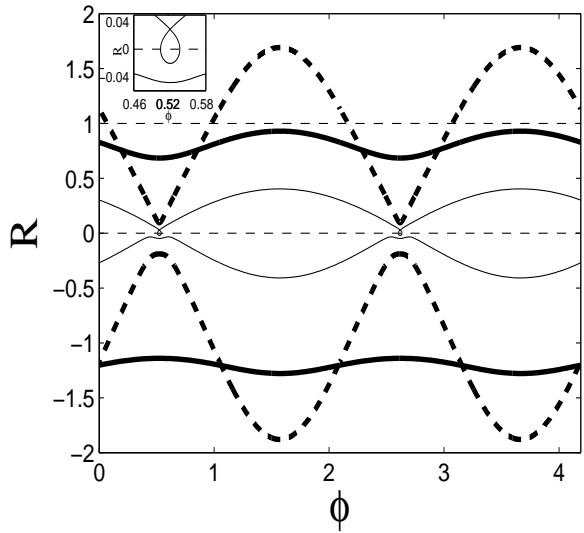}
\end{minipage}
\caption {\label{fig:fig10} Residue curves for the four periodic
orbits with period $2\pi$ indicated by crosses and circles on
Fig.~\ref{fig:fig9} and the two periodic orbits indicated by small
crosses with period $4\pi$ for Case $(II)$. The bold solid curves
are for the upper set of elliptic/hyperbolic orbits with period
$2\pi$. The thin solid curves are for the lower set of
elliptic/hyperbolic orbits with period $2\pi$. The dashed curves
are for the two initially hyperbolic periodic orbits with period
$4\pi$. The inset shows irregular behavior of residue at
$\phi\simeq \pi/6$. The same behavior also occurs at $\phi\simeq
5\pi/6$ for that residue curve.}
\end{figure}

Figure~\ref{fig:fig10} shows the four residue curves of period
$2\pi$ orbits and the two residue curves of period $4\pi$ orbits.
For short periodic orbits (period $2\pi$), the elliptic periodic
orbits remain elliptic and the hyperbolic ones remain hyperbolic
for all values of $\phi$. No bifurcation occurs except in a small
range of phase (see inset of Fig.~\ref{fig:fig10}) where no
significant stability change is observed. Consequently, the
ionization probability is expected to be approximately independent
of $\phi$ and to be lower than Case $(I)$ since for these values
of amplitudes, the chaotic region is smaller. This is consistent
with the experimental and quantum calculations of
Ref.~\cite{PMKoch}. The experimental results show a nearly flat
curve for the ionization probability versus $\phi$, whereas the
quantum calculations show significant variations for this
probability but no sharp increase and decrease as in Case $(I)$.

When the duration of pulse is longer we should take into account
the effect of longer periodic orbits whose residues are shown by
bold dashed curves in Fig.~\ref{fig:fig10} with period $4\pi$ in
this particular case. Apparently the upper period $4\pi$ orbit
experiences period halving bifurcations at $\phi\approx 0.07$ and
$\phi\approx 2.164$; and period doubling bifurcations at
$\phi\approx 0.977$ and $\phi\approx 3.072$.

\begin{figure}
 \centering
 \includegraphics[width=8.5cm,height=7.5cm]{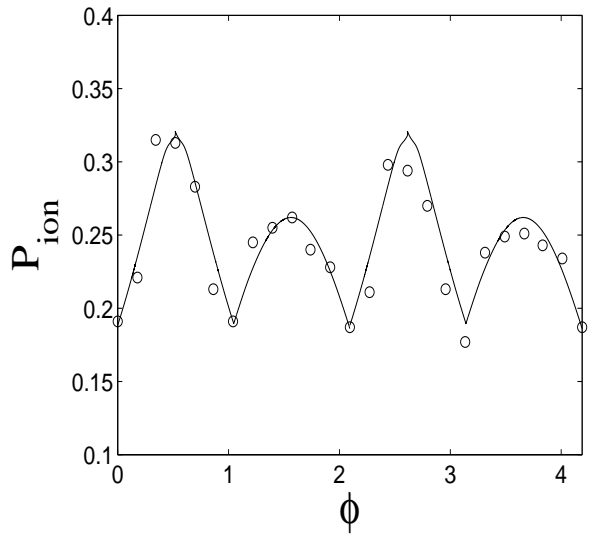}
 \caption{\label{fig:fig11} Normalized ionization probability {\em vs} $\phi$ based on
Eq.~(\ref{Ionempirical}) for Case $(II)$ with $A=-0.485$ and
$B=0.17$. Circles represent the data obtained by one-dimensional
quantum calculations, taken from Ref.~\cite{PMKoch}. Only periodic
orbits with period $2\pi$ are considered.}
\end{figure}

Figure~\ref{fig:fig11} depicts $P_{\mathrm{ion}}$ given by
Eq.~(\ref{Ionempirical}) versus parameter $\phi$ as well as the
data taken from Ref.~\cite{PMKoch} for Case $(II)$. Only periodic
orbits with period $2\pi$ are considered. In
Eq.~(\ref{Ionempirical}) we again take $\phi_{0}=0$ since the
ionization is minimal at $\phi=0$ for Case $(II)$ \cite{PMKoch}.
We notice that it captures some essential features of the
ionization curve, like the two unequal-sized peaks and the
specific shape of both peaks (one more peaked, the other one, more
round). This feature results from the asymmetry property of
bichromatic microwave amplitude \cite{Schafer}. For this case, the
bichromatic microwave field at $\phi=\pi/6\pm\bigtriangleup\phi$
has exactly the same amplitude but opposite direction as at
$\phi=\pi/2\pm\bigtriangleup\phi$, where
$0<\bigtriangleup\phi<\pi/6$. Replacing ``+" in front of the
bichromatic field by ``-" in Eq.~(\ref{Hatom}) leads to a
Hamiltonian for the other direction. Because of this symmetry
property of bichromatic field, the ionization probability along
the other direction can be obtained from Fig.~\ref{fig:fig11} by a
horizontal translation of $\pi/3$. The total ionization rate
should be the sum of the rates for both directions, which shows
equally high peaks for total ionization rate in $\phi$. For longer
pulse duration, we should consider longer periodic orbits like
those with period $4\pi$. Again we should take the baseline
(minimal ionization point) at $\phi=0$. Since in
Fig.~\ref{fig:fig10} the residue value of elliptic periodic orbit
with period $4\pi$ (upper dashed curve) at minimal ionization
point ($\phi=0$) is already greater than $1$, the ionization rate
should be high and roughly identical on the whole $\phi$ space.
This indicates that when pulse duration gets longer, the
ionization rate for any $\phi$ goes higher and suggests ionization
rate for any $\phi$ will reach a roughly constant upper limit if
the time duration is sufficiently long. This property is obviously
true because when the system is exposed to external microwave
field for a longer time, more electrons will gain sufficient
energy to escape. However, as mentioned in Sec.~\ref{sec3}, the
shortest periodic orbits (period $2\pi$) always play the most
dominant role on ionizations rates, while the longer periodic
orbits might be taken into account heuristically for a finer
analysis when the pulse duration gets a lot longer. Finally, our
analysis also indicates the asymmetric ionization property does
not exist for 3:1 mode locking case, in agreement with
Ref.~\cite{Schafer}.

\section{Generalization to $h$:$l$ mode locking}
\label{sec5}

In order to generalize this approach, we also investigate
$h$:$l$=2:1, denoted Case $(III)$, and 5:1, denoted Case $(IV)$
mode locking.

\subsection{Residue curves}
\label{sec5A} 
\begin{figure}
 \centering
 \includegraphics[width=8.5cm,height=7.5cm]{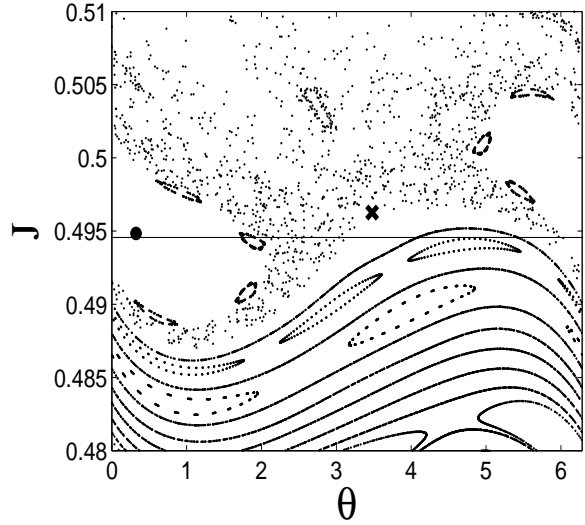}
\caption{\label{fig:fig12}Poincar\'e section of
Hamiltonian~(\ref{HatomAA}) for Case $(III)$ at $\phi=0$. Full
circle (respectively cross) indicates the elliptic (resp.\
hyperbolic) periodic orbit with period $2\pi$ we consider. The
horizontal line corresponds to the principal quantum number
$n=51$.}
\end{figure}

\begin{figure}
 \centering
 \includegraphics[width=8.5cm,height=7.5cm]{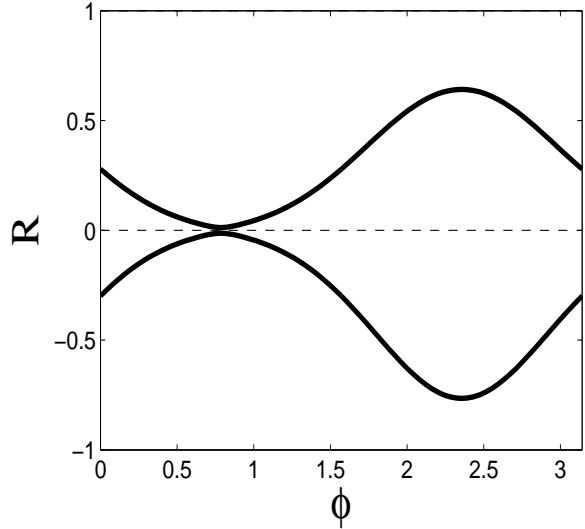}
 \caption{\label{fig:fig13} Residue curves for the two
periodic orbits with period $2\pi$ indicated by cross and circle
on Fig.~\ref{fig:fig12} for Case $(III)$.}
\end{figure}

\begin{figure}
 \centering
 \includegraphics[width=8.5cm,height=7.5cm]{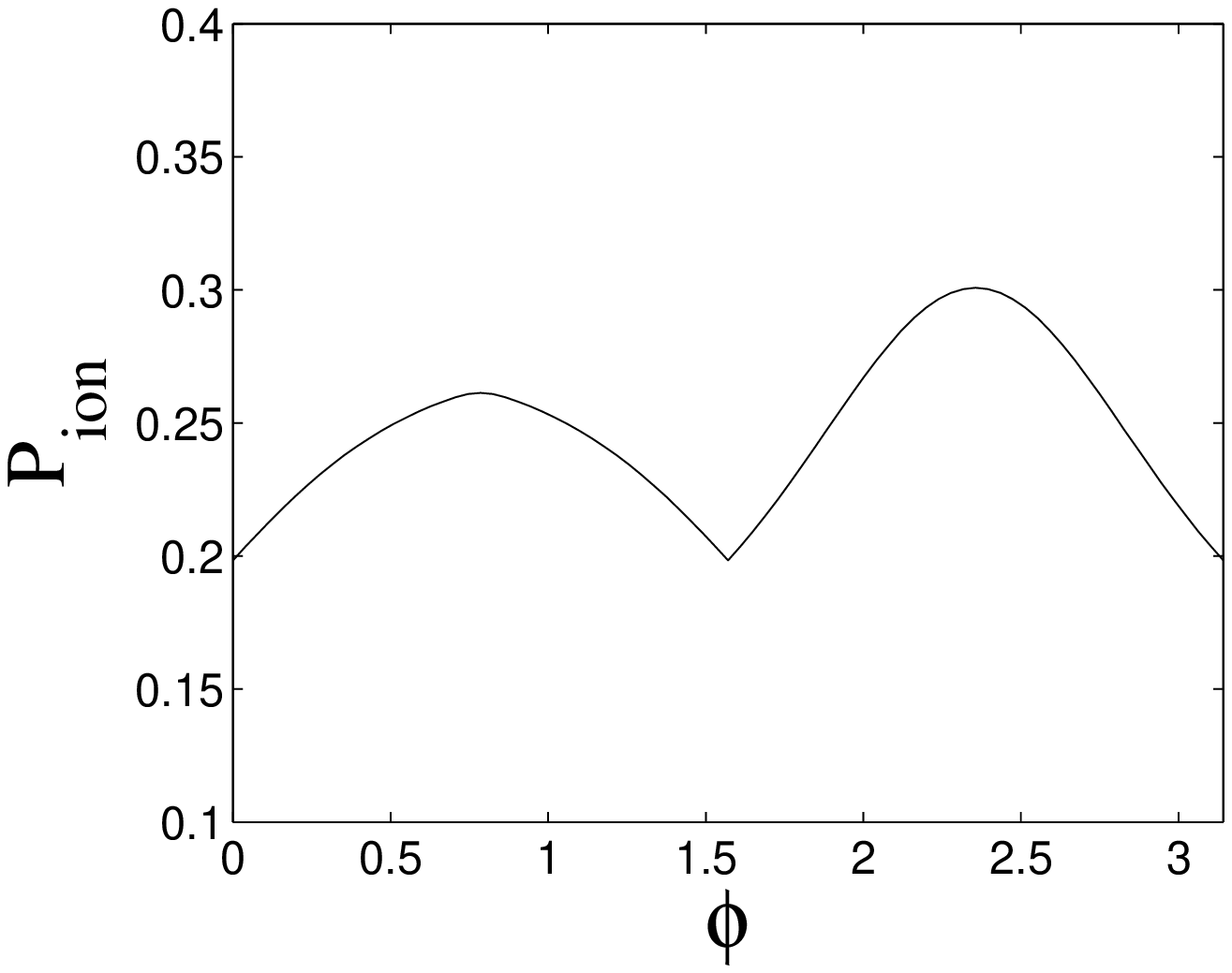}
 \caption{\label{fig:fig14} Relative ionization probability {\em vs} $\phi$ at principal quantum
number $n=51$ based on Eq.~(\ref{Ionempirical}) for Case $(III)$
with $A=0$ and $B=0.1$. The absolute ionization rate may differ
according to the pulse duration.}
\end{figure}

For Case $(III)$, we take $F_h=24\mbox{ Vcm}^{-1}$,
$F_l=53.4\mbox{ Vcm}^{-1}$ and the high frequency of $12\mbox{
GHz}$. Figure~\ref{fig:fig12} shows Poincar\'e section at
$\phi=0$. It indicates the relevant periodic orbits to consider in
the residue analysis. Figure~\ref{fig:fig13} shows the residue
curves based on the two periodic orbits with period $2\pi$ in the
chaotic sea of Fig.~\ref{fig:fig12}. No bifurcations take place
for these orbits. Figure~\ref{fig:fig14} depicts relative
ionization rate with respect to $\phi$ based on
Eq.~(\ref{Ionempirical}). For this case we take $\phi_{0}=0$ in
Eq.~(\ref{Ionempirical}) since the minimal ionization point is at
$\phi=0$ according to the maximum field rule (see next section).
Based on our approach, the asymmetric ionization property along a
single direction (unequal-sized peaks) does appear for the $2:1$
mode locking, in agreement with Ref.~\cite{Schafer}.

\begin{figure}
 \centering
 \includegraphics[width=8.5cm,height=7.5cm]{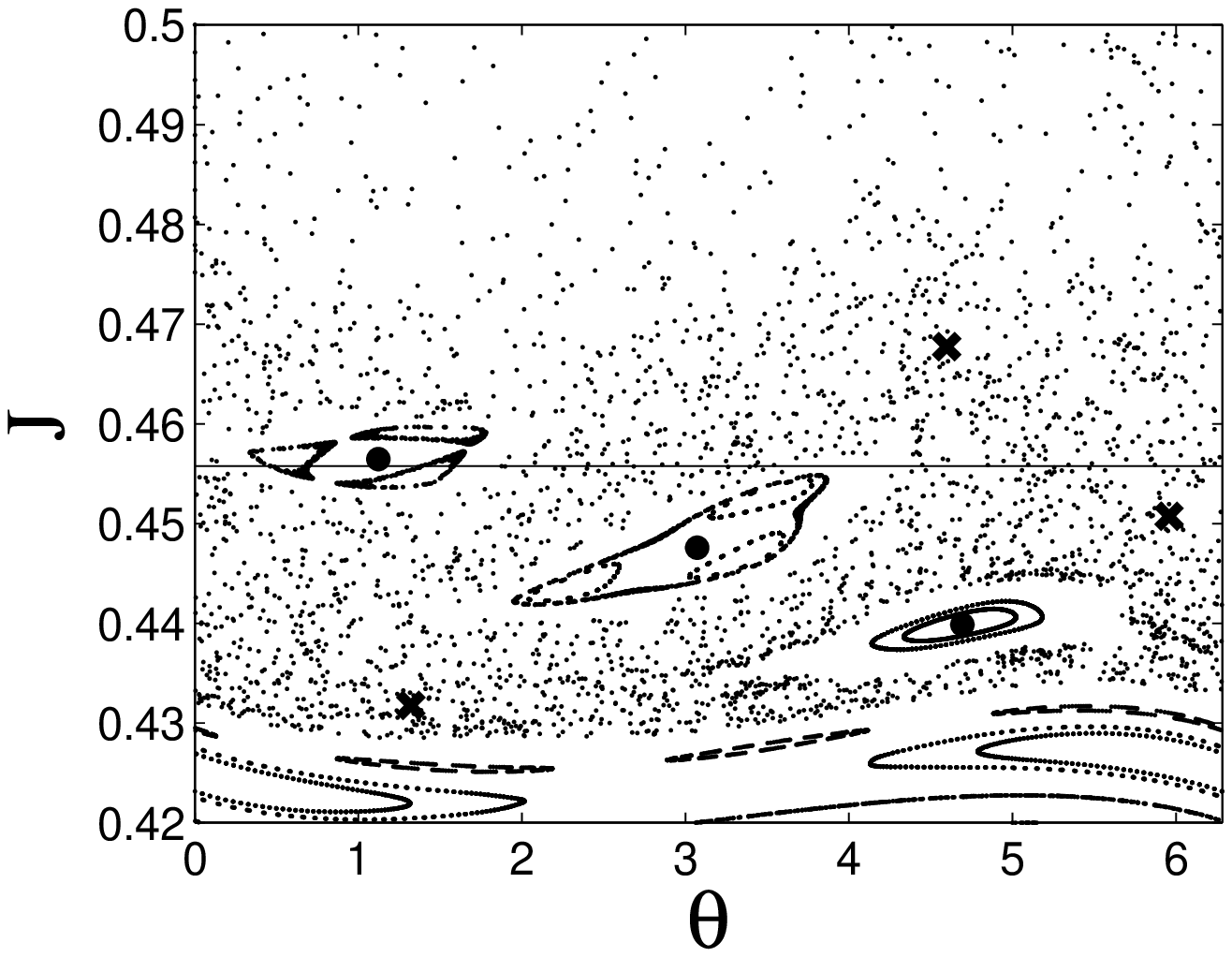}
\caption{\label{fig:fig15}Poincar\'e section of
Hamiltonian~(\ref{HatomAA}) for Case $(IV)$ at $\phi=0$. Full
circles (respectively crosses) indicate the elliptic (resp.\
hyperbolic) periodic orbits with period $2\pi$ we consider. The
horizontal line corresponds to the principal quantum number
$n=47$.}
\end{figure}

\begin{figure}
 \centering
 \includegraphics[width=8.5cm,height=7.5cm]{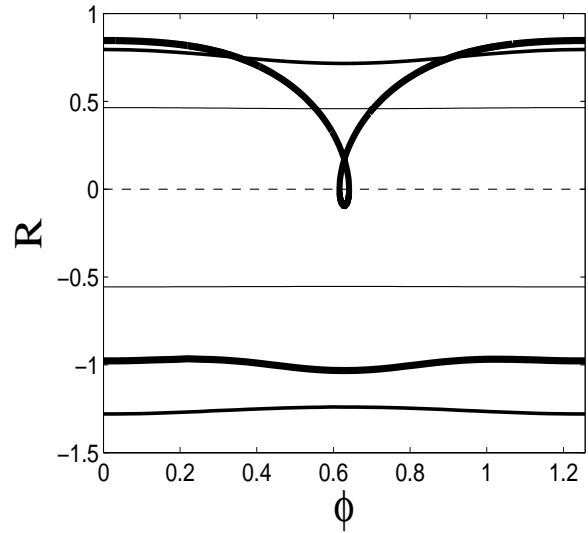}
 \caption{\label{fig:fig16} Residue curves of the six
periodic orbits with period $2\pi$ indicated by crosses and
circles on Fig.~\ref{fig:fig15} for Case $(IV)$.}
\end{figure}

\begin{figure}
 \centering
 \includegraphics[width=8.5cm,height=7.5cm]{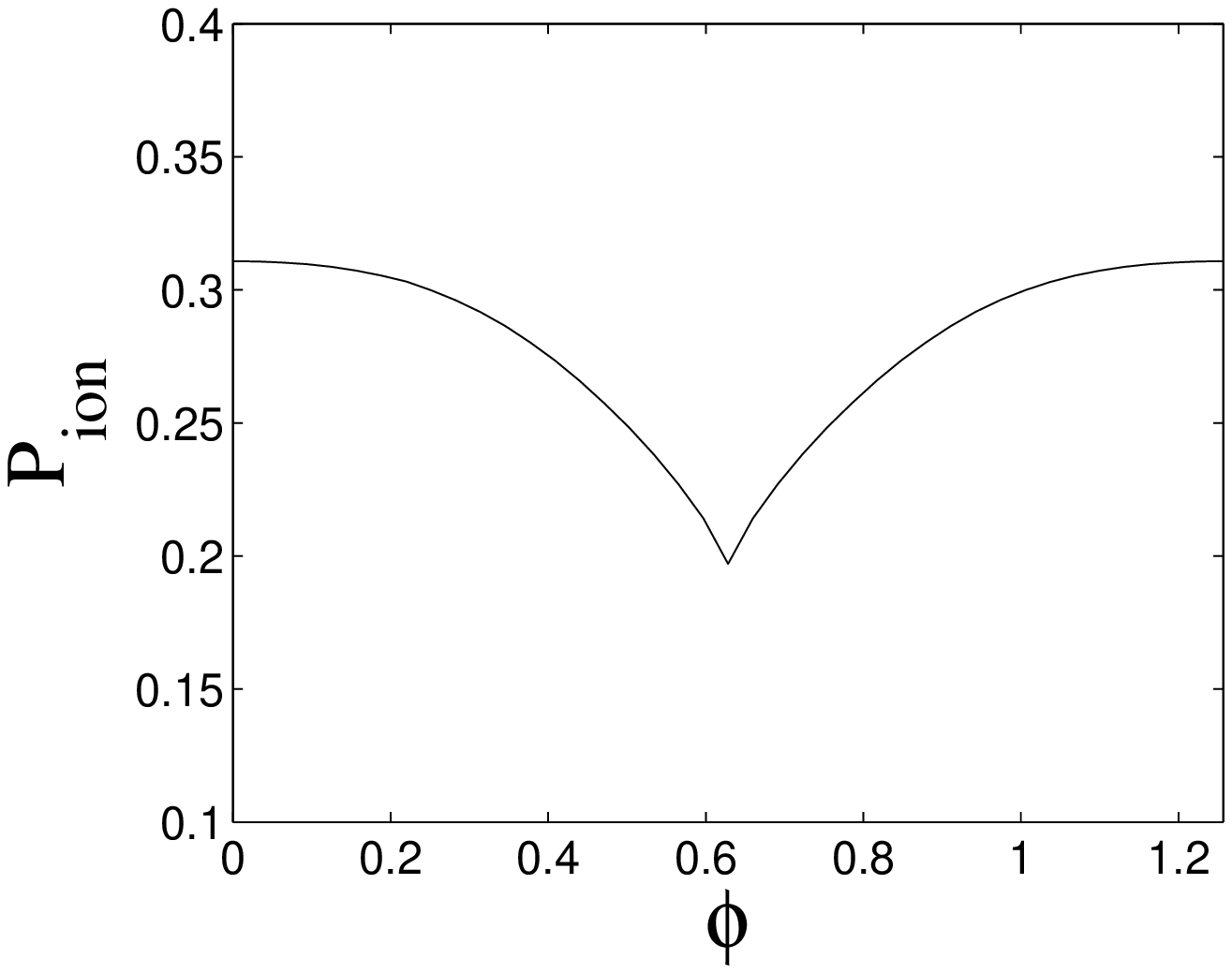}
 \caption{\label{fig:fig17} Relative ionization probability {\em vs} $\phi$ at principal quantum
number $n=47$ based on Eq.~(\ref{Ionempirical}) according to
Fig.~\ref{fig:fig16} for Case $(IV)$ with $A=-0.2$ and $B=0.1$.
The absolute ionization rate may differ according to the pulse
duration.}
\end{figure}

For Case $(IV)$, we take $F_h=24\mbox{ Vcm}^{-1}$, $F_l=53.4\mbox{
Vcm}^{-1}$ and the high frequency is $30\mbox{ GHz}$.
Figure~\ref{fig:fig15} shows a Poincar\'e section at $\phi=0$, and
Fig.~\ref{fig:fig16} shows the residue curves based on the six
periodic orbits of period $2\pi$ in the chaotic sea of
Fig.~\ref{fig:fig15}. We notice that the residue based on the
upper elliptic periodic orbit has similar behavior to the one of
the lower elliptic periodic orbit for Case $(II)$ as shown in the
inset of Fig.~\ref{fig:fig10}. The two residue curves (two thin
lines) in Fig.~\ref{fig:fig16} from the two lower periodic orbits
in Fig.~\ref{fig:fig15} are almost constant and therefore have
little influence on ionization probability. Figure~\ref{fig:fig17}
depicts relative ionization rate with respect to $\phi$ based on
Eq.~(\ref{Ionempirical}) according to Fig.~\ref{fig:fig16} (The
lowest two periodic orbits are not considered since their
corresponding residue curves are almost constant). For this case
we take $\phi_{0}=\pi/5$ in Eq.~(\ref{Ionempirical}) since the
minimal ionization point is at $\phi=\pi/5$ according to the
maximum field rule. The asymmetry ionization property does not
exist for the 5:1 mode locking.

\subsection{Comparison with the maximum field rule}
\label{sec5B}

\begin{figure}
 \begin{minipage}[t]{8cm}
 \centering
 \includegraphics[width=6.cm,height=5.cm]{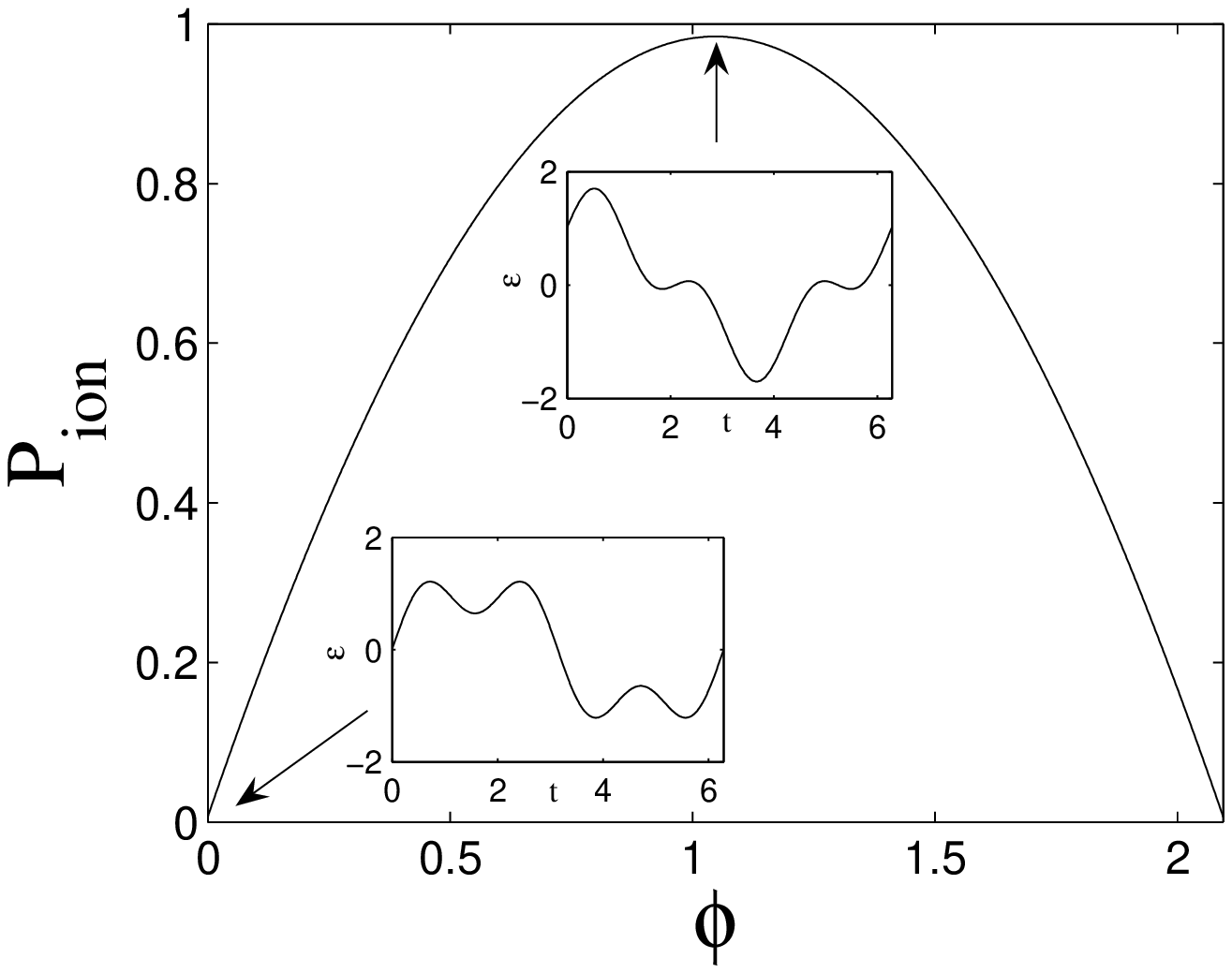}
 \mbox{{\bf (a)} }
 \end{minipage}
 \centering
 \begin{minipage}[t]{8cm}
 \includegraphics[width=6.cm,height=5.cm]{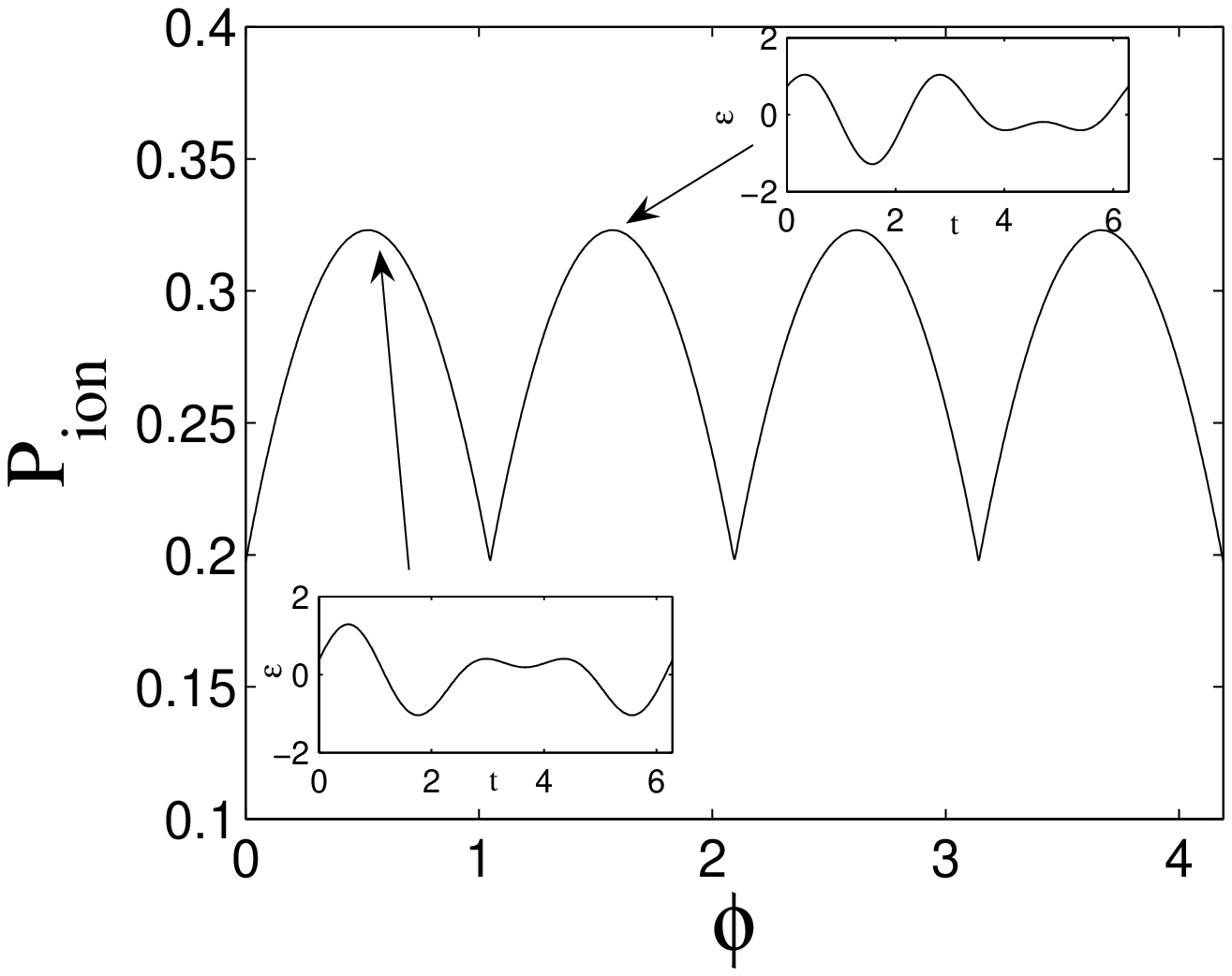}
 \mbox{{\bf (b)} }
 \end{minipage}
 \begin{minipage}[t]{8cm}
 \centering
 \includegraphics[width=6.cm,height=5.cm]{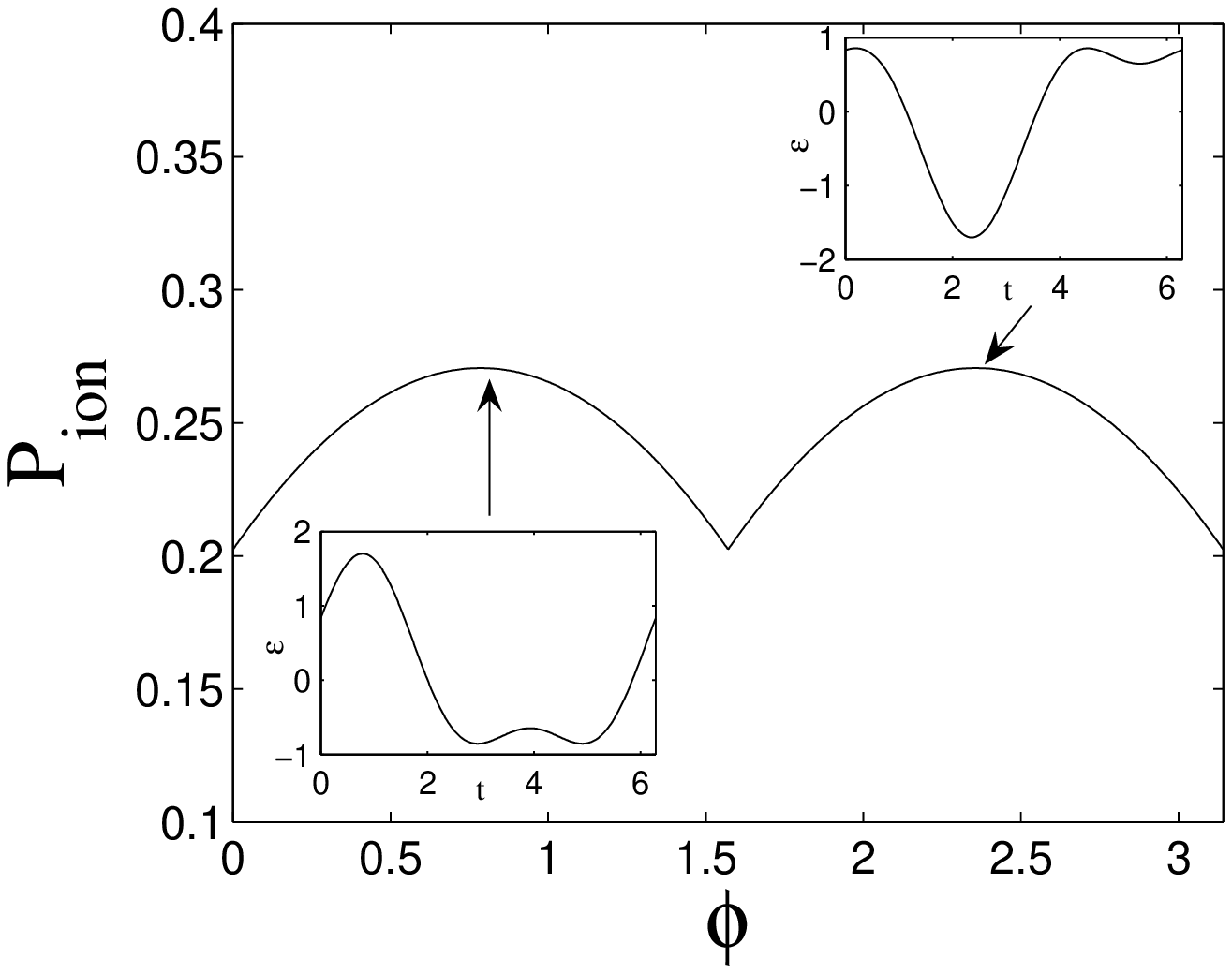}
 \mbox{{\bf (c)} }
 \end{minipage}
 \centering
 \begin{minipage}[t]{8cm}
 \includegraphics[width=6.cm,height=5.cm]{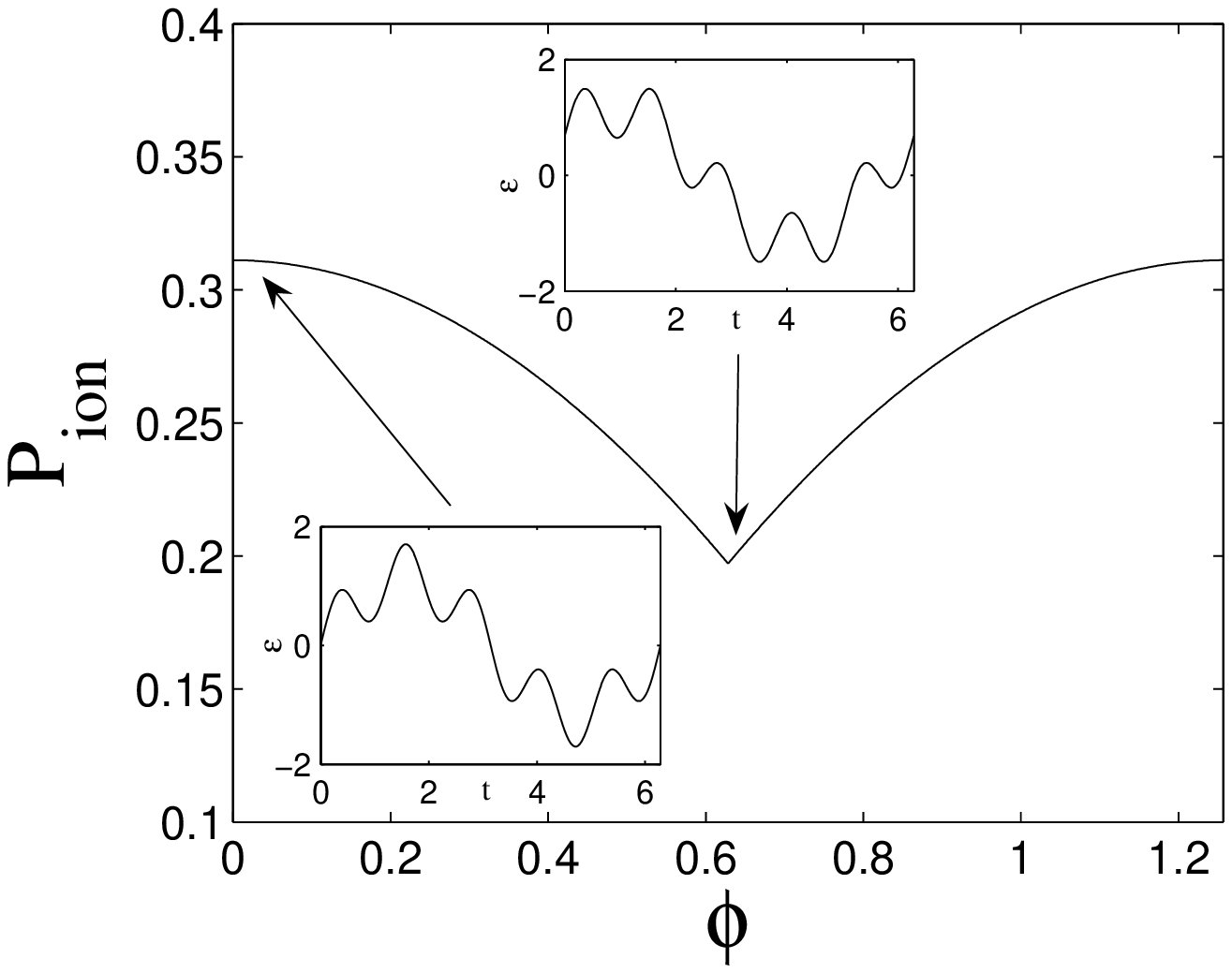}
 \mbox{{\bf (d)} }
 \end{minipage}
\caption{\label{fig:fig18} Relative ionization probability based
on the maximum field rule, on Eq.~(\ref{peakamplitude}), for $(a)$
Case $(I)$ with $C=-2.42$ and $D=2$. Insets show the bichromatic
field for $\phi=0$ (bottom panel) and $\phi=\pi/3$ (top
panel).$(b)$ Case $(II)$ with $C=-2.25$ and $D=2$. Insets show the
bichromatic field for $\phi=\pi/6$ (bottom panel) and $\phi=\pi/2$
(top panel). $(c)$ Case $(III)$ with $C=-0.24$ and $D=0.3$. Insets
show the bichromatic field for $\phi=\pi/4$ (bottom panel) and
$\phi=3\pi/4$ (top panel). and $(d)$ Case $(IV)$ with $C=-0.625$
and $D=0.55$. Insets show the bichromatic field for $\phi=0$
(bottom panel) and $\phi=\pi/5$ (top panel). Arrows indicate the
$\phi$ values for which the bichromatic fields are drawn.}
\end{figure}

A traditional way of explaining ionization probability behavior is
the maximum field rule (peak field amplitude rule) formulated in
Ref.~\cite{Sirko,Sauer}. The relative ionization probability {\em
vs} $\phi$ based on the maximum field rule can be written in the
form
\begin{equation}
\label{peakamplitude} P_{\mathrm{ion}}(\phi)=C+D
\mathrm{max}_{t\in[0,2\pi]}|F_{h}\sin(ht)+F_{l}\sin (l t+\phi)|.
\end{equation}
The parameters $C$ and $D$ in Eq.~(\ref{peakamplitude}) are merely
a translation and a dilatation of the curve in order to make the
curves from the maximum field rule comparable to those based on
bifurcation analysis. Figure~\ref{fig:fig18} shows relative
ionization probability based on Eq.~(\ref{peakamplitude}) for four
different cases. Compared with our periodic orbit bifurcation
analysis in finite pulse duration, apparently the maximum field
rule does not result in quantitative agreement with the quantum
simulation results of Ref.~\cite{PMKoch} for either Case $(I)$ or
$(II)$. In Case $(III)$, the maximum field rule does not produce
the two unequal-sized peaks as our bifurcation analysis does. Only
in Case $(IV)$ do the results based on the maximum field rule
agree with our bifurcation analysis because there is neither
bifurcation nor asymmetric ionization property for this case.
Insets of Fig.~\ref{fig:fig18} show one cycle ($t\in[0, 2\pi]$)
bichromatic field $\epsilon(t)= F_{h} \sin(h t)+F_{l} \sin (l
t+\phi) $ with typical parameters as used for our previous
analyses in dimensionless units for different $\phi$ values.
Arrows indicate the $\phi$ values for which the bichromatic fields
are drawn. Generally for Case $(I)$ the positive component of the
bichromatic field is the same as the negative component for any
$\phi$ in each full cycle (directional symmetry), whereas this
directional symmetry does not appear for Case $(III)$
\cite{Schafer} except for some specific $\phi$ values like
$\phi=0, \frac{\pi}{2}, \pi$. Similarly, for Case $(II)$ this
directional symmetry generally does not appear except for some
specific $\phi$ values like $\phi=0, \frac{\pi}{3},
\frac{2\pi}{3}, \pi, \frac{4\pi}{3}$, whereas for Case $(IV)$,
this directional symmetry does appear for any $\phi$ values.
Although the maximum field rule can be used to determine
qualitative features like the minimal ionization point with
respect to $\phi$, the quantitative agreement is not as
satisfactory as the one given by a method which relies on
analyzing the chaotic dynamics like the one used in this article
based on periodic orbit bifurcation analysis to depict relative
ionization.

\section*{Acknowledgements}
This research was supported by the US National Science Foundation.
C.C. acknowledges support from Euratom-CEA (contract
EUR~344-88-1~FUA~F). We thank Xavier Leoncini and Luca Perotti for
useful discussions.

\end{document}